\documentclass[conference]{IEEEtran}
\IEEEoverridecommandlockouts
\usepackage{cite}
\usepackage{amsmath,amssymb,amsfonts}
\usepackage{algorithmic}
\usepackage{graphicx}
\usepackage{textcomp}
\usepackage{xcolor}
\usepackage{authblk}
\usepackage{hyperref}
\usepackage{float}
\usepackage{chngcntr}
\usepackage{array}
\usepackage{caption}
\DeclareCaptionFont{myfont}{\normalfont}
\def\BibTeX{{\rm B\kern-.05em{\sc i\kern-.025em b}\kern-.08em
    T\kern-.1667em\lower.7ex\hbox{E}\kern-.125emX}}
\begin{document}

\title{An Investigation of Software Defined Wide Area Networking (SD-WAN) for Optimizing Multi-site
Enterprise Networks \\
% {\footnotesize \textsuperscript{*}Note: Sub-titles are not captured in Xplore and
% should not be used}
% \thanks{Identify applicable funding agency here. If none, delete this.}
}

\author[ ]{%
  \begin{minipage}[t]{0.45\textwidth}
    \centering
    Chaoran Sun \\
    School of Computer Science \\
    University of Nottingham, England \\
    \textit{alycs10@Nottingham.ac.uk}
  \end{minipage}%
  \hspace{0.1\textwidth} 
  \begin{minipage}[t]{0.45\textwidth}
    \centering
    Milena Radenkovic \\
    School of Computer Science \\
    University of Nottingham, England \\
    \textit{milena.radenkovic@nottingham.ac.uk}
  \end{minipage}%
}

\maketitle

\begin{abstract}
\textbf{
Enterprise networks are becoming increasingly complex, posing challenges for traditional WANs in terms of scalability, management, and operational costs. Software Defined Networking (SDN) and its application in Wide Area Networks (SD-WAN) offer solutions by decoupling the control plane from the data plane, providing centralized management, enhanced flexibility, and automated provisioning. This research investigates the challenging application of SD-WAN to optimize traditional multi-site enterprise networks. Experimental scenarios are designed in which SD-WAN is implemented on a traditional multi-site network topology with complex architecture, then followed by comprehensive evaluations of its performance across various critical aspects, including hardware status, transmission performance, and security.
}
\end{abstract}

\section{Introduction}
To comply with technological trends and meet the expansion needs of enterprises themselves, enterprise networks are becoming increasingly large and complex. Multi-site enterprises usually use Wide Area Networks (WANs) for communication because of the large geographical spans between sites. However, the huge WAN architecture, coupled with the relatively static and rigid nature of traditional networks, makes these enterprises lack flexibility when expanding their scale and introducing new features. In addition, the increase in network traffic brought about by business growth makes the management and maintenance of the WAN itself more difficult. Traditional enterprise networking is facing challenges in terms of scalability, management difficulty, and operation and maintenance costs. Therefore, many of today's multi-site enterprises require more flexible, scalable and efficient network solutions.

Software Defined Networking (SDN) emerges as a transformative technology that addresses these challenges by decoupling the control plane from the data plane, offering centralized control and management, enhanced flexibility, and automated provisioning. The combination of SDN and WAN, Software Defined Wide Area Networking (SD-WAN) has become a new solution to address the above challenges. 

The research aims to explore the application SD-WAN to optimize traditional geographically dispersed enterprise networks to help them better adapt to today's increasingly large and complex network structures. The research will design an experimental scenario which implements SD-WAN architecture based on a traditional multi-site enterprise networking topology, evaluate the impact of SD-WAN on traditional networks by implementing the automated configuration and centralized management functions of SD-WAN.

This research designs a multi-site enterprise network scenario through a network simulator, and then introduces SD-WAN technology to transform some areas of the network to explore the optimization of SD-WAN architecture on traditional network operation and maintenance. Section 2 discusses some related concepts and previous research. Section 3 shows the design of the scenario, including the design topology diagrams of the enterprise network and the transformation plan of applying SD-WAN. Section 4 provides a detailed description of the experiments design. Section 5 evaluates the overall performance of SD-WAN and  compares the transmission efficiency of traditional and optimized networks. Section 6 summarizes the experimental results and discusses the limitations of the experiment and future research.

\section{Background and Related Work}
With the expansion of data centers and the rise of cloud computing, traditional network architectures are increasingly unable to cope with dynamic changes and rapid expansion. In traditional network architecture, the control plane and data plane of network devices (such as routers and switches) are tightly integrated. To deal with these challenges, the concept of Software Defined Networking (SDN) is introduced, which separates network control from data forwarding, thereby making the network more flexible and programmable. For example, it has been found that while SDN introduces performance trade-offs, it simplifies network management and supports advanced functionalities like dynamic routing and centralized control \cite{gelberger2013performance}. In addition, a study introduces an adaptive, predictive resource provisioning model that integrates distributed analytics and machine learning to manage resources dynamically. It focuses on edge cloud services to enhance responsiveness and efficiency in distributed environments \cite{huynh2021distributed}. This is instructive for exploring the function of SDN technology in enhancing network dynamics in this project. SDN technology has evolved rapidly, from initial theoretical research to practical application, and gradually expanded to a wider range of network environments, especially enterprise networks.

Software Defined Wide Area Networking (SD-WAN) is a specific application that based on SDN technology concepts and focuses on optimizing and managing wide area network connections. SD-WAN can be defined as a technology that optimizes the operations over WANs with cloud-based management and automation. It separates data control plane to allow centralized controllers and to manage the control plane of the overlay network \cite{rajagopalan2020overview}. Many researchers divide the structure of SD-WAN into three layers, as shown in Figure 2.1. The control layer and data layer are the core layers of SD-WAN, which directly affect the performance and function realization of the network. This kind of multi-layer architecture is becoming mainstream in recent years, such as the MODiToNeS platform in the mobile network field, this modular design allows the network to respond to changes more quickly and flexibly \cite{radenkovic2016towards}.

\setcounter{figure}{0} 
\renewcommand{\thefigure}{2.1} 
\begin{figure}
    \centering
    \includegraphics[width=0.75\linewidth]{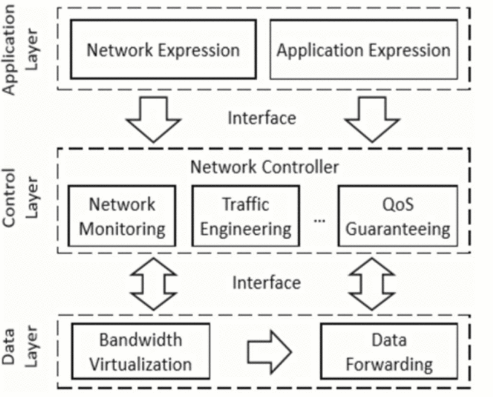}
    \caption{SD-WAN: Logical Architecture
    \cite{yalda2022survey}}
    \label{fig:enter-label}
\end{figure}

The development background of SD-WAN stems from the expansion needs of enterprise WANs. Traditional WANs typically rely on dedicated lines such as MPLS to connect geographically dispersed branch offices. However, as enterprises globalize and cloud computing applications increase, traditional WANs are costly and inflexible. The emergence of SD-WAN aims to dynamically optimize traffic management, reduce costs and improve application performance through a software-defined approach. Therefore, SD-WAN is considered a promising architecture for next-generation wide area networks. \cite{yang2019software}. 

Many previous studies have assessed SD-WAN's impact on network performance, focusing mainly on dynamic routing and centralized management under various configurations. For example, A study examines various SD-WAN architectures, highlighting their benefits in improving network performance and reducing costs through intelligent path selection and centralized management \cite{segevc2020sd}. Another study presents an open-source approach to implementing SD-WAN, providing insights into the practical aspects of deploying and managing SD-WAN solutions in enterprise settings \cite{troia2020sd}. In addition to optimizing network performance, SD-WAN also has advantages in compatibility with modern technologies.It is found that the application of deep reinforcement learning can optimize traffic management in SD-WAN environments, indicating that SD-WAN can also introduce advanced machine learning techniques to enhance network adaptability and performance \cite{troia2020deep}. A recent study introduces a novel On-POP-Overlay architecture that can significantly enhance interconnectivity, latency, and security of multi-site Software-Defined Data Centers (SDDC). This architecture optimizes network paths and introduces potential for AI-driven route selection in SD-WAN frameworks, showing substantial improvements in performance through empirical testing across global data centers \cite{wang2024sd}.

\section{Scenario Design}

In an enterprise WAN, infrastructure is typically deployed by segmenting the network into different areas, such as headquarter, data center and branches, each area is responsible for different functions. The headquarter is the core of the enterprise network and is responsible for centralized management and control of the entire enterprise's IT infrastructure and network policies. The data center is responsible for storing and processing the enterprise's core data and key applications, it also provides IT services to the headquarters and branch offices via WAN. Branches are the frontline responsible for specific business execution and provide localized services and support. Each of these areas performs their own duties, allowing enterprises to operate efficiently and ensuring that resources in various geographical locations can be seamlessly connected and work together through the WAN.

This research first designed a typical traditional multi-branch enterprise network scenario, as illustrated in the topology diagram in Figure 3.1. The topology contains headquarter, data center and 4 branches. In additional, to emphasize the cross-regional distribution of the enterprise network, the scenario sets up 2 service providers to simulate cross-regional WAN connections. In the scenario, all areas use the BGP protocol for external communication and each area is distinguished by an independent BGP AS number, while internal communication within the area is established through EIGRP or OSPF protocols. 

\renewcommand{\thefigure}{3.1} 
\begin{figure}
    \centering
    \includegraphics[width=0.75\linewidth]{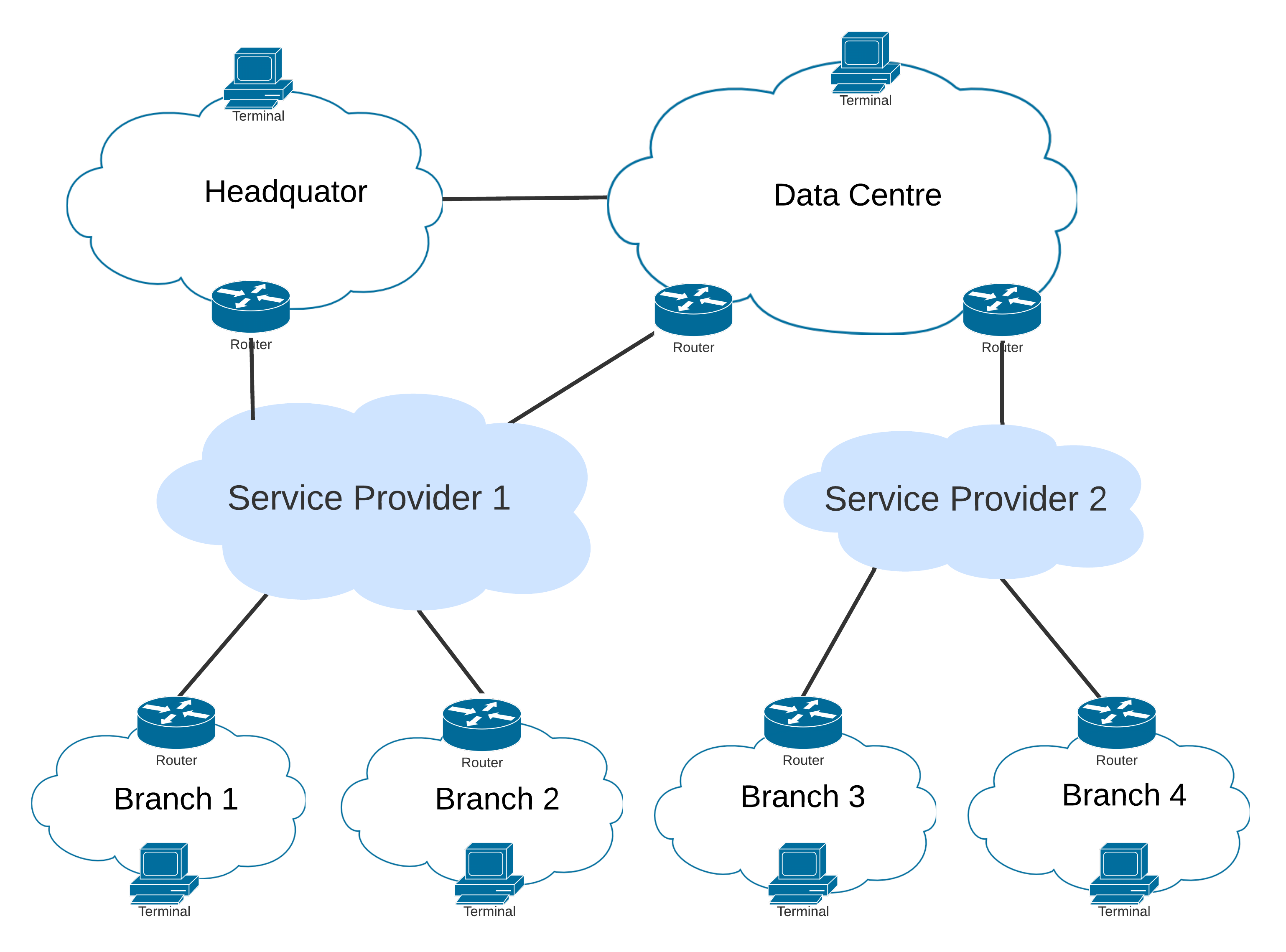}
    \caption{Traditional Enterprise Network Topology}
    \label{fig:enter-label}
\end{figure}

As today's enterprise networking requirements for flexibility and scalability increase, the shortcomings of traditional networks have become increasingly apparent. Traditional WANs rely on hardware-based routers, manually configure routing policies. The configuration of network devices will become more difficult if network devices keep increasing, and the cost of configuration and management will also increase \cite{nomura1999policy}. To solve these problems, the research introduced SD-WAN to experimentally transform the data centre area and part of branches based on the original traditional network, as illustrated in the topology diagram in Figure 3.2.

\renewcommand{\thefigure}{3.2} 
\begin{figure}
    \centering
    \includegraphics[width=0.75\linewidth]{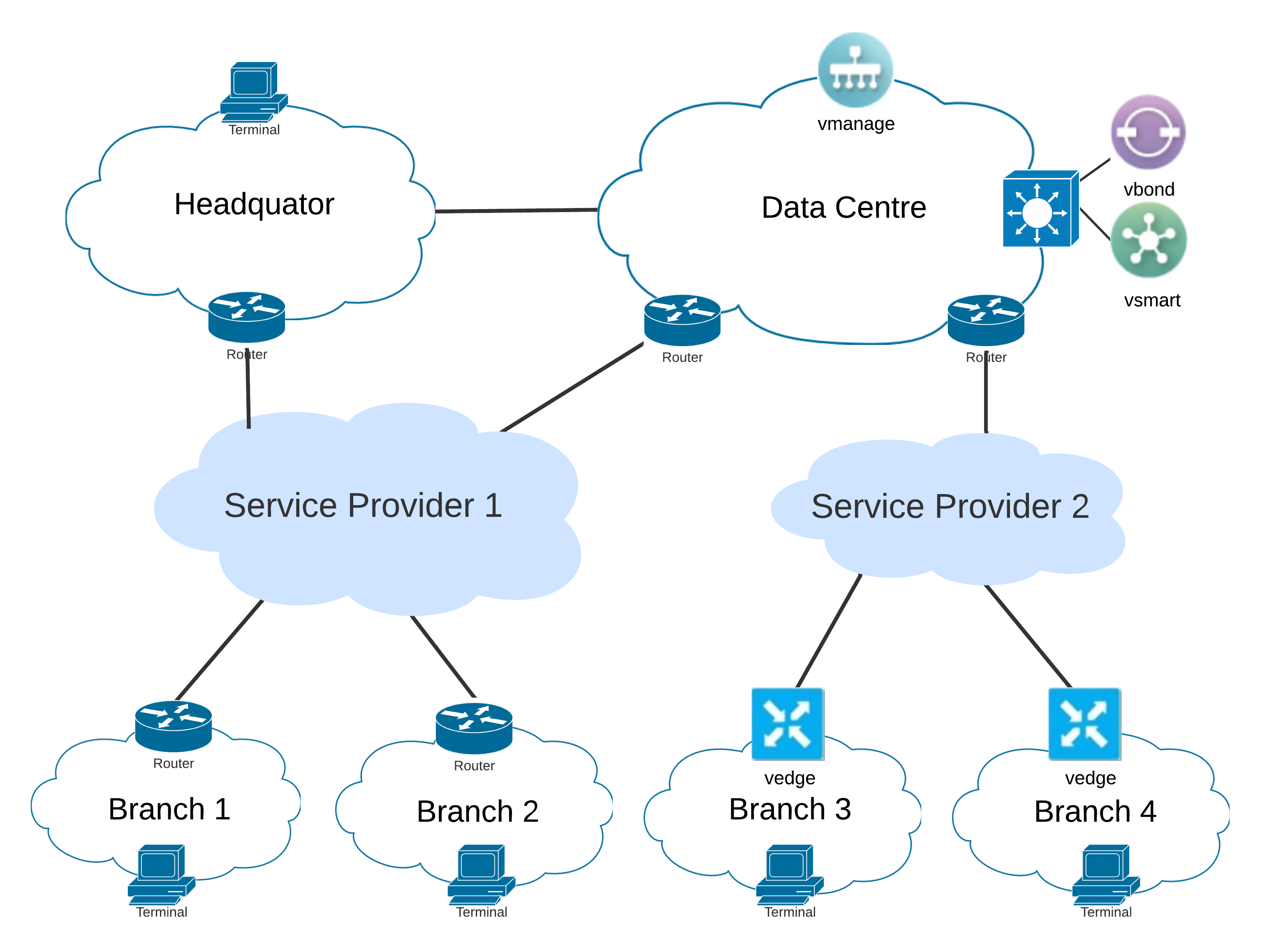}
    \caption{SD-WAN Network Topology}
    \label{fig:enter-label}
\end{figure}

The SD-WAN architecture consists of 4 core omponents: vManage, vBond, vSmart and vEdge. These components form a complete SD-WAN solution that provides centralized management, high flexibility and secure network connectivity. The vManage serves as management controller, it provides a centralized management platform for SD-WAN solutions. As coordinating controller, the vBond responsible for device authentication and initial connection in the SD-WAN network. When devices join the SD-WAN network, vBond ensures that the devices can discover each other and establish secure connections. The vSmart is the policy controller responsible for managing and distributing routing policies and security policies. The vEdges are the edge routers and responsible for actual data forwarding. Through SD-WAN's centralized control panel, enterprises can manage and configure the entire network from one location, reducing the need for on-site configuration and reducing operation and maintenance costs and complexity. 

For topology optimization, SD-WAN controller components vManage, vBond and vSmart are deployed in the data center. In branches 3 and 4, traditional routers are replaced by vEdge devices. For most modern enterprise networking, data centers usually host an enterprise's core applications and services, SD-WAN deployment in the data center can reduce the delay of management traffic and control signaling, and improve network performance and application response speed. By deploying SD-WAN in the data center, a more reliable control center can be provided for the headquarters, branches and other sites. Especially when the headquarters and data center back up each other, it can enhance the redundancy and resilience of the entire network.

\section{Experiments Design}
\subsection{Experiments Preparation}
The simulator used for experiments is EVE-NG. EVE-NG is a comprehensive network emulation platform that allows users to create and configure virtual network environments to simulate real-world networking scenarios. EVE-NG allows the flexible adjustment of network devices and their parameters. This simulator will be used to create virtual network environments that simulate real-world multi-site enterprise networks.

The experiments utilize Cisco Viptela SD-WAN virtual device images to establish the 4 key SD-WAN components: vManage, vBond, vSmart and vEdge. As one of SD-WAN solutions, Cisco SD-WAN based on Viptela has been considered to be the preferred one for large, complex deployments \cite{wang2019software}, which makes it suitable for enterprise networking. The conventional switches and routers in the experiments use Cisco's IOL virtual device image, as efficient simulation tools with low resource usage, they are ideal for use in simulation environments. The experiments import these device images into the simulator to build network infrastructure and realize the functions of SD-WAN in this scenario. The host device image uses the version that comes with EVE-NG, and does not need to be imported from outside. The information of the devices used in the experiments are as Table 4.1.

\begin{table}
    \renewcommand{\thetable}{4.1}
    \centering
    \scalebox{1}{
    \begin{tabular}{|l|l|l|}
        \hline
        Devices & Type & Version \\ \hline
        vManage, vBond, vSmart, vEdges & Cisco Viptela & Viptela 18.4.4 \\ \hline
        Routers & Cisco vIOL & 15.2 \\ \hline
        Switches & Cisco vIOL-L2 & 15.7(3)M2 \\ \hline
        Hosts & VPCS & 1.3(0.8.1) \\ \hline
    \end{tabular}}
    \caption{Devices Type and Image Version}
    \label{tab:my_label}
\end{table}

\subsection{Underlying Network Design}
Due to the use of a network simulator, this research focuses on the design of a logical structure to reflect the geographical distribution characteristics of each region of a multi-site enterprise. The detailed topology diagram is shown in the figure 4.1.

\renewcommand{\thefigure}{4.1} 
\begin{figure}
    \centering
    \includegraphics[width=0.75\linewidth]{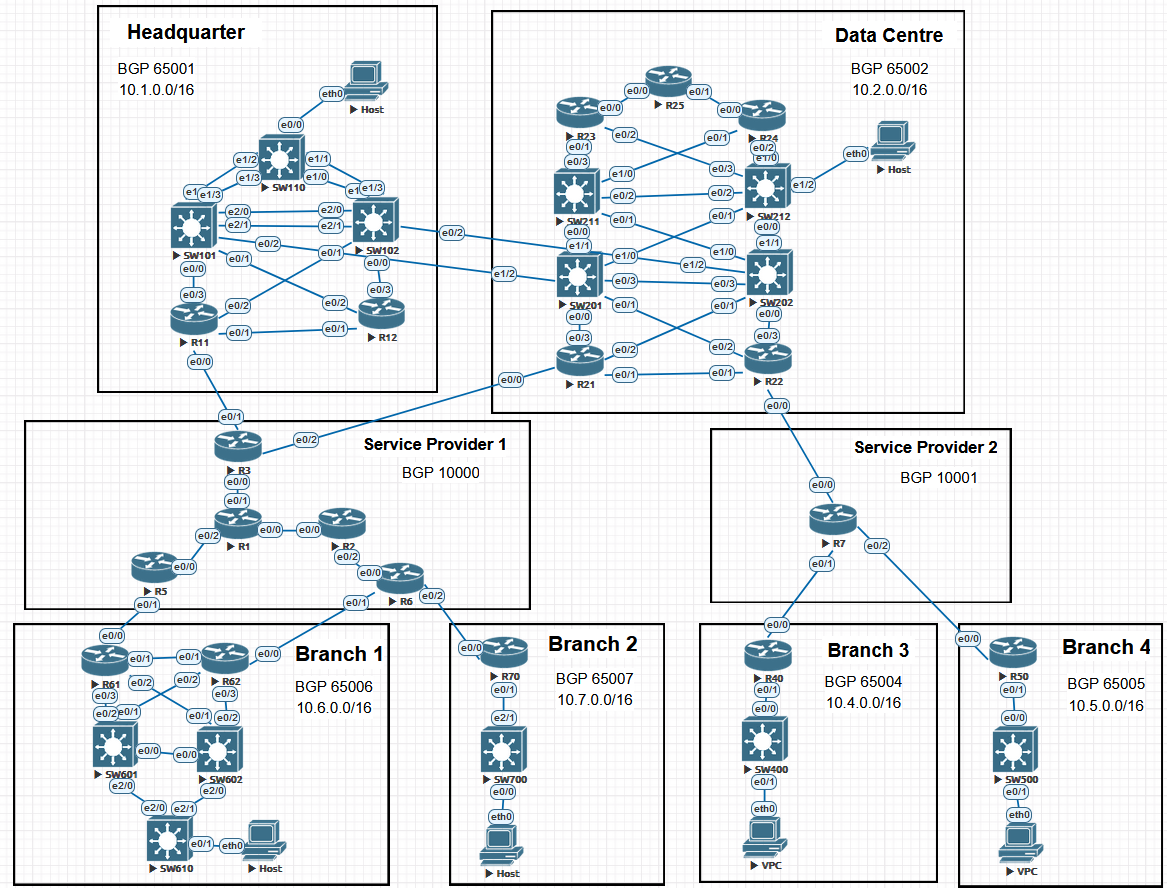}
    \caption{Underlying Network Architecture Topology Diagram}
    \label{fig:enter-label}
\end{figure}

In the headquarter and data center, each network area has multiple core switches and routers connected to each other to form a redundant triangle or ring structure. This design ensures that even if a device or link fails, the network can continue to work through other paths. In addition to cross-region communication through the service provider, a direct dedicated line is deployed between the headquarters and the data center to enable quick access to critical applications and real-time data synchronization.

To reflect the scale difference between the two service providers, the typologies distinguishes their scales by the number of routers. Service Provider 1 (SP1) has a total of 6 routers, while Service Provider 2 (SP2) is represented by a single router. Branches 1 and 2 are connected to SP1, while branches 3 and 4 are connected to SP2. This design balances the load and improves the overall stability of the network.

For protocol selection, BGP is a path vector form of distance vector routing protocol and capable of handling large numbers of routing entries \cite{huston2010securing}, making it ideal for enterprise-level networks, especially when sites are connected through different service providers. In this scenario, all areas use BGP to communicate, each area is configured as an independent autonomous system (AS), thereby maintaining routing autonomy within each area. After BGP is configured on the border router, it can effectively manage the exchange of routing information inside and outside the enterprise and support complex network environments across multiple autonomous systems, it also enables each area to manage its internal routing independently.

Internal communication in each area uses EIGRP and OSPF. As a hybrid protocol widely used in Cisco devices environments, EIGRP is an advanced variant of the distance  vector  protocol with  several characteristicslike link-state protocols \cite{azman2021configure}. EIGRP is suitable for small and medium-sized enterprise networks and network environments that require simpler configuration and management. OSPF is a link-state routing protocol, it uses logical arranged area to build its topology \cite{rakheja2012performance}. This feature makes OSPF very suitable for managing complex network topology and supports multi-area divisions. Therefore, OSPF is the preferred protocol for building large enterprise networks and service provider networks. EIGRP and OSPF can provide flexible, scalable, and efficient internal routing solutions for enterprise networks, meeting the needs of different regions and departments. Therefore, this experiment combines the two protocols to enhance the realism of the simulation. The network segmentation and routing protocols configured for each area are as Table 4.2.

\begin{table}
    \renewcommand{\thetable}{4.2}
    \centering
    \scalebox{0.95}{
    \begin{tabular}{|c|c|c|c|}
        \hline
        Area & BGP AS Number & Network Segment & Internal Protocols \\ \hline
        Headquarter & 65001 & 10.1.0.0/16 & EIGRP \& OSPF \\ \hline
        Data Centre & 65002 & 10.2.0.0/16 & EIGRP \& OSPF \\ \hline
        Branch 1 & 65006 & 10.6.0.0/16 & EIGRP \\ \hline
        Branch 2 & 65007 & 10.7.0.0/16 & EIGRP \\ \hline
        Branch 3 & 65004 & 10.4.0.0/16 & OSPF \\ \hline
        Branch 4 & 65005 & 10.5.0.0/16 & OSPF \\ \hline
    \end{tabular}}
    \caption{Network Segmentation and Routing Protocols by Area}
    \label{tab:my_label}
\end{table}

In configuration methods, manual configuration is the preferred way to configure network devices in traditional IT environments. Therefore, all routers and switches adopt the traditional manual configuration method when establishing traditional enterprise network.

\subsection{SD-WAN Architecture Design}
The design of SD-WAN involves the construction of a multi-layer architecture. In the SD-WAN architecture, traditional networking is usually regarded as an underlay network (data layer). The underlay network refers to the physical network or infrastructure layer, which provides the basic path and connection for data packet transmission. This not only refers to the various areas of the enterprise network, but also includes the service providers they span. SD-WAN uses virtualization and abstraction technologies to build a logical overlay network layer (control layer) on top of the underlay, and implements advanced functions such as intelligent routing, traffic optimization, encryption, security policies, etc. through centralized management platform. Therefore, SD-WAN devices need to undergo an onboarding process before they are officially put into use.

The topology diagram of the deployed SD-WAN architecture is shown in Figure 4.2. The deployment scheme of SD-WAN is as follows: The SD-WAN controllers (vManage, vBond and vSmart) are deployed in the data center area, and the vManage replaces the original host as the terminal for centralized management. These three devices only need to be configured with basic host information and static routes for them to communicate with each other. The edge device components (vEdges) are deployed in the Branch 3 and Branch 4 areas and replace the original border routers. In each branch, the vEdges are manually configured only with the IP address of the physical port and a static route to the data center; no other initial configuration is performed. The goal is to ensure that these Vedges can successfully ping the SD-WAN controller components in the data centre. All other configurations on vEdges will be completed through automatic configuration in the next steps.
\renewcommand{\thefigure}{4.2} 
\begin{figure}
    \centering
    \includegraphics[width=0.75\linewidth]{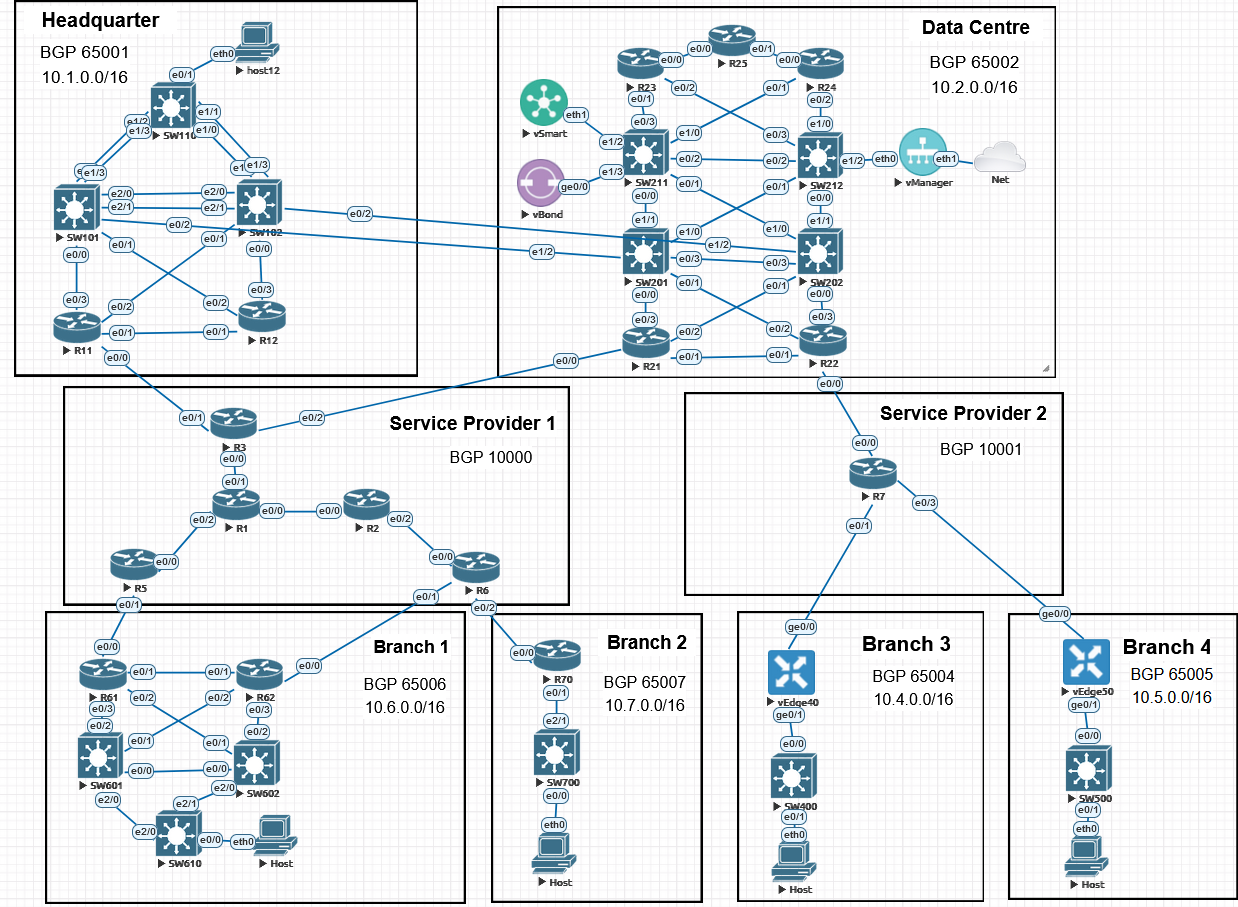}
    \caption{SD-WAN Architecture Topology Diagram}
    \label{fig:enter-label}
\end{figure}

For onboarding process, After completing the underlying communication of each SD-WAN component, access vManage and perform onboarding process for SD-WAN related devices to build the overlay network. Firstly, generate and import the certificates of vBond and vSmart through the vManage command line. After authentication, vManage will perform automatic synchronization to connect the 2 devices into the SD-WAN. When synchronization is complete, check the running status of the two devices through the vManage graphical management platform, as shown in Figure 4.3. The access of vBond will enable the management platform to obtain the currently available vEdge device serial numbers. Therefore, the next step is to select the available serial numbers to import the vEdge devices that need to be connected to SD-WAN, then the management platform will automatically perform the same synchronization process as before. Finally, all components are connected to SD-WAN, these online devices can be seen in the main interface of the management platform, which means the onboarding process of the SD-WAN components is completed.
\renewcommand{\thefigure}{4.3} 
\begin{figure}
    \centering
    \includegraphics[width=0.75\linewidth]{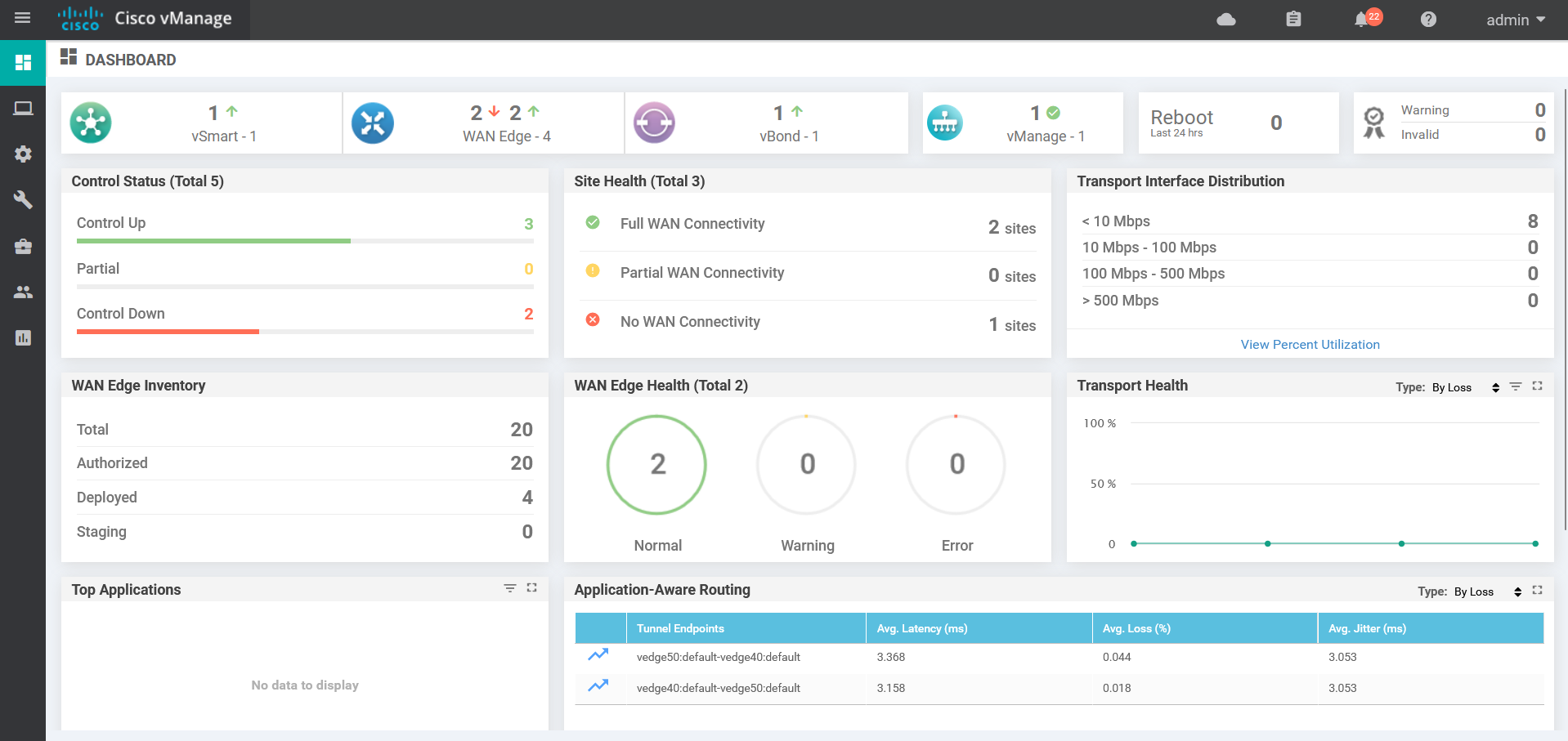}
    \caption{SD-WAN Graphical Management Platform}
    \label{fig:enter-label}
\end{figure}

The management platform provides real-time network status monitoring and displays comprehensive information in a centralized manner, including control status, site health status, transmission status, etc. This visual representation enables administrators to quickly identify problems in the network, such as devices going offline, connectivity being lost, or performance being degraded.

\subsection{SD-WAN Configuration Template Design}
SD-WAN achieves automated configuration through the template function. This experiment implements remote configuration from the data center to the branches by designing configuration templates on the management platform and pushing them to the vEdge devices. 

A template consists of several features, which define the configuration and operation behavior of the device. A template consists of several features, which define the configuration and operation behavior of the device. There are several common types of features: System features, which are used to define the basic settings of the device, such as the host name, system number, etc. Interface features are used to configure the physical or virtual interfaces associated with a specific VPN. Routing features are used to define the routing settings in the VPN, such as static routing, and the configuration of dynamic routing protocols such as BGP or OSPF. By flexibly combining these features, administrators can create customized configuration templates to suit different business requirements and network architectures. 

In this experiment, the configurations of the original border routers in Branch 3 and Branch 4 are designed into templates, and then sent to the vEdge devices to make them serve as new border routers. The features used for branch 3 and branch 4 for creating configuration templates are as Figure 4.4.

\renewcommand{\thefigure}{4.4} 
\begin{figure}
    \centering
    \includegraphics[width=0.75\linewidth]{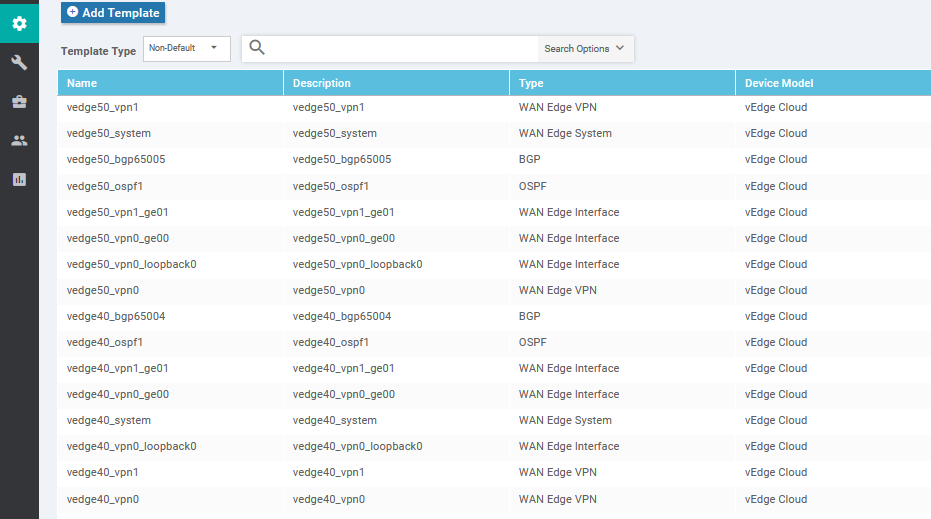}
    \caption{Features for Creating Configuration Templates}
    \label{fig:enter-label}
\end{figure}

When the template design is completed, they are converted into executable configuration by the vSmart, and then pushed to the vEdge devices respectively. All configuration processes in this stage are performed on the management platform, and no other configuration is required for the vEdge device manually. If the configuration needs to be modified, the operator only needs to modify the template on the management platform and push it again.

\section{Evaluation}

This experiment collected various data to evaluate the performance of SD-WAN in many critical aspects, including hardware status, transmission performance and security.

\subsection{Hardware Status}
The management platform provides monitoring of the hardware status of all vEdge, vSmart, vBond and vManage devices in the SD-WAN network. Taking Figure 5.1 as an example, the management platform has a dashboard dedicated to monitoring hardware information. Although hardware conditions in a simulated environment are not as critical as in a real-world setting, monitoring some basic metrics can provide a more intuitive understanding of the running status of SD-WAN devices. For devices in the simulation environment, monitoring the CPU and memory can help ensure that the system resource utilization is within a reasonable range to prevent system crashes caused by resource exhaustion. 

\renewcommand{\thefigure}{5.1} 
\begin{figure}
    \centering
    \includegraphics[width=0.75\linewidth]{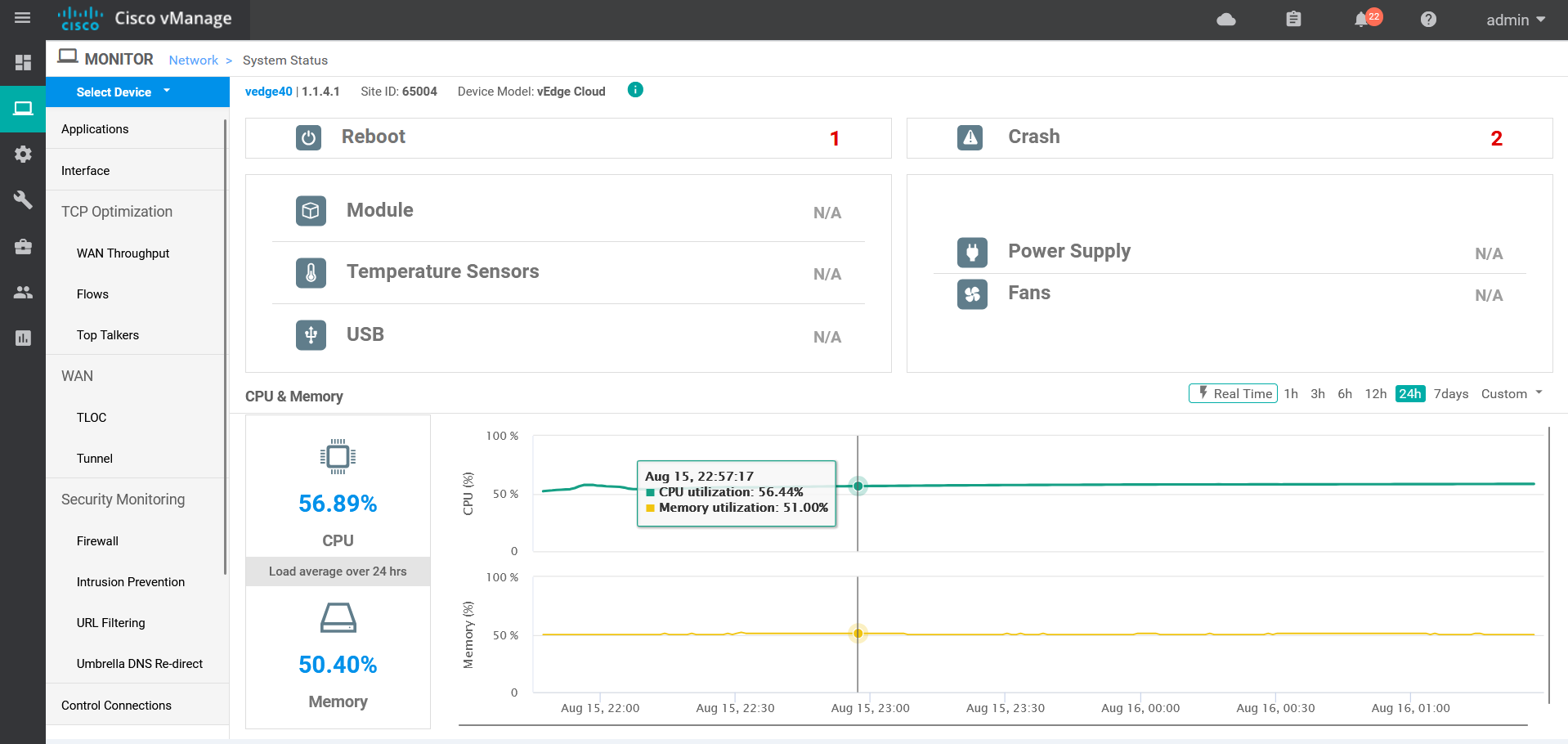}
    \caption{Hardware Information Dashboard}
    \label{fig:enter-label}
\end{figure}

When the SD-WAN architecture is in a stable operating state, the overall fluctuation of hardware values is minimal, so this experiment collects real-time data in a specific time period. The CPU and memory data of all SD-WAN devices collected from management platform dashboard are as Table 5.1.

\begin{table}[h!]
\renewcommand{\thetable}{5.1}
\centering
\scalebox{0.8}{
\begin{tabular}{|l|c|c|c|c|}
\hline
\textbf{Device Name} & \textbf{Number of CPUs} & \textbf{Memory Total} & \textbf{CPU Usage} & \textbf{Memory Usage} \\ \hline
vManage  & 4  & 16384 MB & 8.17\%  & 55.53\%  \\ \hline
vBond    & 2  & 2048 MB & 5.14\%  & 59.60\%  \\ \hline
vSmart   & 2  & 2048 MB & 0.47\%  & 19.15\%  \\ \hline
vEdge 40 & 2  & 2048 MB & 56.89\% & 50.40\%  \\ \hline
vEdge 50 & 2  & 2048 MB & 55.67\% & 49.93\%  \\ \hline
\end{tabular}}
\caption{Device CPU and Memory Usage}
\label{tab:cpu_memory_usage}
\end{table}

According to the recorded data, The CPU usage of these devices is generally below 60\%. Notably, the CPU usage of the SD-WAN controllers (vManage, vBond and vSmart) is below 10\%. In addition to the extremely low memory usage of vSmart (below 20\%), memory usage of other devices are generally in the range of 50-60\%. Overall, the resource utilization of the SD-WAN devices under the current configuration is stable and normal. Additionally, there is sufficient capacity available to accommodate additional load or future expansion, ensuring the devices can handle increased demands without compromising performance.

\subsection{Transmission Performance}

This research also includes a controlled experiment designed to evaluate the routing functionality and transmission efficiency after introducing SD-WAN into the enterprise network. The experiment simulates cross-region access within the enterprise network. The experimental setup involved comparing the two topologies: one representing an enterprise network integrated with SD-WAN architecture, and the other representing a traditional architecture (the original topology). Data packets were sent from a host in the headquarters to another host (or vManage) in the data center. The same number of packets was then sent from the host (or vManage) in the data center to the border routers (or vEdges) in Branch 3 and Branch 4. Finally, route reachability was verified, and data such as Round-Trip Time (RTT) were collected. 

The transmitted packets were ICMP packets, which are ideal for conveying control information and error reporting based on the Internet Control Message Protocol. If the destination is reachable, the packets are returned. In each test, 100 packets of the same size were sent. The statistical data from these tests are presented in Tables 5.2 and 5.3. Based on the data in the tables, a bar chart comparing TTL was also created, as shown in Figure 5.2.

\begin{table}[h!]
\renewcommand{\thetable}{5.2}
\centering
\renewcommand{\arraystretch}{1.5}
\scalebox{0.66}{
\begin{tabular}{|l|l|c|c|c|c|c|}
\hline
\textbf{Source} & \textbf{Destination} & \textbf{Packet Size} & \textbf{TTL} & \textbf{Max RTT} & \textbf{Min RTT} & \textbf{Avg RTT} \\ \hline
Headquarter (Host) & Data Centre (Host) & 84 bytes & 61 & 9.077 ms & 2.131 ms & 3.53638 ms\\ \hline
Data Centre (Host) & Branch 3 (R40) & 84 bytes & 61 & 8.524 ms & 1.812 ms & 3.010 ms \\ \hline
Data Centre (Host) & Branch 4 (R50) & 84 bytes & 61 & 8.296 ms & 2.003 ms & 2.943 ms \\ \hline
\end{tabular}
}
\caption{Transmission Performance in Traditional Architecture}
\label{tab:host_rtt}
\end{table}

\begin{table}[h]
\renewcommand{\thetable}{5.3}
\centering
\renewcommand{\arraystretch}{1.5}
\scalebox{0.645}{
\begin{tabular}{|l|l|c|c|c|c|c|}
\hline
\textbf{Source} & \textbf{Destination} & \textbf{Packet Size} & \textbf{TTL} & \textbf{Max RTT} & \textbf{Min RTT} & \textbf{Avg RTT} \\ \hline
Headquarter (Host) & Data Centre (vManage) & 84 bytes & 61 & 31.315 ms & 3.343 ms & 8.030 ms \\ \hline
Data Centre (vManage) & Branch 3 (vEdge 40) & 84 bytes & 60 & 24.540 ms & 2.957 ms & 6.640 ms \\ \hline
Data Centre (vManage) & Branch 4 (vEdge 50) & 84 bytes & 60 & 19.824 ms & 3.146 ms & 7.268 ms \\ \hline
\end{tabular}
}
\caption{Transmission Performance in SD-WAN Architecture}
\label{tab:sdwan_rtt}
\end{table}

\renewcommand{\thefigure}{5.2} 
\begin{figure}
    \centering
    \includegraphics[width=0.75\linewidth]{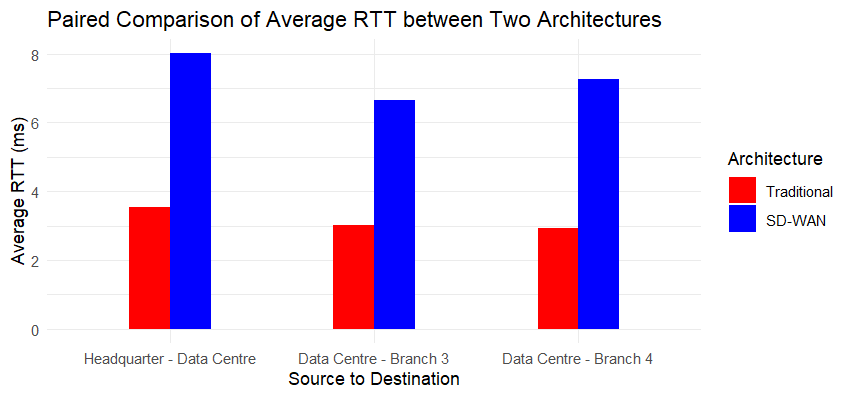}
    \caption{Comparison of Average RTT Between Architectures}
    \label{fig:enter-label}
\end{figure}

From the results, in both the traditional and SD-WAN architectures, all target destinations were successfully reached by the data packets. This indicates that the SD-WAN's routing functionality correctly routes packets from the source to the destination. The presence of measurable RTT data further confirms the connectivity of the paths and the reachability of the packets.

In the traditional architecture, the RTT values are lower and more consistent, indicating that in a simpler network topology, data transmission delays are minimal and the path stability is high. In the SD-WAN architecture, RTT values increase across all paths, with significant increases observed in both maximum and average RTT. Compared with traditional network architecture, SD-WAN's multi-layer structure enables more dynamic data flow, which brings higher transmission latency.

The possible reason for this phenomenon is that in the simulation environment, the delay may be relatively uniform and the packet loss rate is relatively fixed, and there will be no obvious dynamic changes. This is different from the characteristics of large physical distance and high dynamics of enterprise networking in reality.

\subsection{Security}

SD-WAN has an integrity assurance mechanism that uniformly distributes configurations through templates to ensure that each device strictly follows predefined security policies and avoid configuration deviations due to human errors or malicious operations. To verify this mechanism, this experiment attempts to rewrite the current configuration directly through the vEdge command line after the configuration template is sent to the vEdge device. As a result, no changes can be written to the vEdge configuration, and the vEdge CLI is locked as "read-only", which means SD-WAN successfully managed the permissions of the device.
Centrally manage configuration through the vManage control platform, concentrating configuration rights in the hands of a few administrators, reducing the risk of configuration errors and unauthorized configuration changes. Restricting direct CLI access helps enforce the principle of least privilege and prevents unauthorized configuration changes. In this scenario, this mechanism increases the security of the border router and effectively protects the enterprise from internal threats that may come from branches and other areas. 

Compared to traditional network architectures, where threats are more dispersed across various network elements, in SD-WAN, security threats tend to be more centralized. This centralization makes it easier to detect abnormal behavior in network traffic that may be caused by an attacker \cite{shu2016security}.

\section{Conclusion}\
This research investigates the application of SD-WAN in multi-site enterprise networks and its optimization effect on network performance through simulation experiment. The experiment designs a typical multi-site enterprise network topology, introduces and deploys the SD-WAN architecture based on it, and then evaluates the performance of SD-WAN in terms of hardware status, transmission performance, and security.

The experimental results show that SD-WAN significantly improves the management and configuration efficiency of enterprise networks. SD-WAN's centralized management platform can effectively monitor and manage network status, reduce the risks and cost caused by manual configuration, and provide higher network resilience. From a performance perspective, while the SD-WAN's multi-layer architecture may offer more flexible automated configuration and improved network management, it comes at the cost of higher latency. Therefore, enterprises need to balance the benefits of SD-WAN's flexibility against the potential increase in delay when considering its adoption.

Although this research tested SD-WAN performance through simulation, the simulated environment differs from the real enterprise network's complexity, limiting its ability to fully reflect SD-WAN's advantages in path selection and QoS. Therefore, the more advantages and potential of SD-WAN need to be verified in real scenarios.

Overall, the traditional enterprise WANs has a simpler structure, but has limitations in management and optimization, especially in responding to dynamic business needs and improving network resiliency. SD-WAN's optimized enterprise WANs greatly improves the flexibility and efficiency of the network by introducing centralized management and control and automated configuration, while simplifying the management process, allowing enterprises to better adapt to modern and complex IT environments. 

Future research should focus on more complex test environments to fully evaluate the performance of SD-WAN in real enterprise networks. The various features of SD-WAN enable it to flexibly integrate with a wide range of technologies, such as programmable networks, artificial intelligence, and mobile networks, and future research will also explore these aspects. In terms of security, there is also room for enhancement. For example, integrating privacy protection mechanisms such as the k-anonymity-based Location Privacy-Aware Forwarding (LPAF) protocol can enhance privacy protection across broader network environments \cite{radenkovic2016cognitive}.

\bibliographystyle{IEEEtran}

\bibliography{references.bib}

\end{document}